\documentstyle[aps,preprint,epsf,citesort]{revtex}

\begin{document}

\title{Topological Defects in Size-Dispersed Solids}

\author{M. Reza Sadr-Lahijany,$^1$ Purusattam Ray,$^{2,1}$
and H. E. Stanley$^1$}

\address{$^1$ Center for Polymer Studies and Department of Physics\\ 
Boston University, Boston, Massachusetts 02215\\ 
$^2$ The Institute of Mathematical Sciences, CIT Campus, Chennai 600
113, India}

\date{submitted to Phys. Rev. Letters 7 Jan 2000}

\maketitle

\begin{abstract}

We study the behavior of the topological defects in the inherent
structures of a two-dimensional binary Lennard-Jones system as the size
dispersity varies. We find that topological defects arising from the
particle size dispersity are responsible for destabilizing the solid as
follows: (i) for particle density $\rho \leq 0.9$, the solid melts
through intermediate states of decreasing hexatic order arising from the
proliferation of unbounded dislocations, (ii) for $\rho > 0.9$, the
dislocations form grain boundaries, dividing the system into
micro-crystallites and destroying the translational and orientational
order.

\end{abstract}

\bigskip\bigskip

%\begin{multicols}{2}

Topological defects play a crucial role in melting of a solid,
especially in two dimensions.  In two dimensions (2D) these defects,
present in bound pairs at low temperature solid phase, are believed to
unbind and destroy the crystalline order as temperature is raised
causing melting~\cite{nelson2}.  Like temperature, size-dispersity
(inhomogeneity in particle size) {\it disfavors\/} crystalline order,
and can even convert a solid to a liquid~\cite{Barrat-Hansen,srs}.  Here
we study the defect morphology in the inherent structures of a 2D
Lennard-Jones system with a bimodal distribution of particle sizes.

We simulate $N=10^4$ particles interacting with a truncated
``shifted-force Lennard-Jones'' pair-potential in
2D.  We choose half the particles to be smaller than the rest, and
define the size dispersity $\Delta$ to be the ratio of the
difference in their sizes to the mean size.  We start by placing the
particles randomly on the sites of a triangular lattice, embedded in a
rectangular box of edges $L_x$ and $L_y$ with aspect ratio
$L_x/L_y=\sqrt 3/2$ (to accommodate close-pack hexagonal structure
without distortion).  We apply periodic boundary condition and use the
velocity Verlet method to integrate Newton's equation of motion. Units
are set by choosing the mass of the particles and the Lennard-Jones (LJ)
energy and length scales to be unity. The density $\rho$ is
the ratio of the area occupied by the particles to the box area.

We equilibrate a state, defined by $(\rho,\Delta)$, at a constant
temperature $T=1$ using Berendsen's thermostat; at this temperature the
2D solid can form for low enough dispersity. We run our simulation at
constant energy until the temperature $T$, the pressure $P$, and the
energy $E$ stabilize with less than 1\% fluctuation (typically for
$2\times 10^5$ time steps, where the time step is 0.01 in LJ units).  We
also check that the average particle displacement is at least a few
times the average particle size. To obtain higher density states, we
increase $\rho$ in steps of 0.01 from 0.85 (liquid state) to 1.05 by
gradually compressing the box, keeping the aspect ratio fixed, and
equilibrating the system at each of these densities.

We perform defect analysis on 100 equilibrated configurations for each
of the state points ($\rho,\Delta$). Clear identification of geometrical
defects is difficult in simulations due to the presence of many
``virtual defects'' which arise because of vibrational excitations.  To
overcome this difficulty, we analyze the inherent structure of each
configuration \cite{stillinger}, obtained by removing the vibrational
excitations---or equivalently by minimizing (locally) the potential
energy using the conjugate gradient method.

In order to find the defects, we construct a Voronoi cell around each
particle, thereby uniquely defining its nearest neighbors. The ordered
close-packed structure of the 2D solid is hexagonal, so each particle
$i$ at position $\mbox{\bf r}_i$ has $n_i=6$ neighbors.  A defect arises
when a particle has $n_i\ne 6$, which generates a ``topological charge''
$q_i\equiv n_i-6$ \cite{clark}.  Defects which are nearest neighbors are
grouped as defect clusters with total charge $Q\equiv\sum_i q_i$ and
total dipole moment $\mbox{\bf P}\equiv\sum_i \mbox{\bf r}_iq_i$, where
the sum is over all defects $i$ in the cluster.  Figure~\ref{defsfig}
shows an example of defects in an otherwise ideal triangular lattice.

We find that the defects fall into three categories: (i) {\it
Monopoles}. Clusters with $Q\neq 0$. The simplest case is a
disclination---a size-one cluster (Fig.~\ref{defsfig}a). (ii) {\it
Dipoles}. Clusters with $\mbox{\bf P}\neq 0$. The simplest case is a
dislocation, a size-two cluster (Figs.~\ref{defsfig}a
and~\ref{defsfig}b), composed of a ``bound pair'' of neighboring defects
of opposite charge.  (iii) {\it Blobs}. Clusters with $Q=\mbox{\bf
P}=0$. The most common case is a quadrupole (Fig.~\ref{defsfig}a) made
of a ``bound pair'' of neighboring dislocations with oppositely-oriented
dipole moments (see, e.g., \cite{nelson3,Rubinstein}).

Figure~\ref{r9defs} shows typical snapshots of the inherent structure at
$\rho=0.9$ for three values of $\Delta$. The corresponding phase points
($\rho,\Delta$) represent solid, hexatic and liquid phases \cite{srs}.
In the low-dispersity solid phase [Figs.~\ref{r9defs}a and
\ref{r9defs}b], defects occur mostly in the form of blobs (quadrupoles).
Such defects, being without charge or dipole moment, cause little
distortion in the nearby order and so are energetically inexpensive. We
find that the defects in the inherent structure of the solid phase
aggregate to form domains separated by nearly defect-free regions,
suggesting that there is an effective attraction between defects. This
effective attraction may arise from packing constraints since in dense
packing, formation of a local large-amplitude defect is improbable.  As
$\Delta$ increases, more defects in the form of dipoles are created.

At large defect density, free dislocations appear.  Figure~\ref{r9defs}c
shows their presence, in the hexatic phase.  A dislocation destroys the
long-range translational order as it introduces an extra half row
(Fig.~\ref{defsfig}b) that can only terminate in another dislocation
with equal but opposite dipole moment.  Translational order is destroyed
over the range of separation of the dislocation pair \cite{nelson2}.
Dislocations, however, retain orientational order. We find in
configurations such as Fig.~\ref{r9defs}c that 2D translational order is
lost (due to the abundance of dipoles). However, orientational order
shows an algebraic decay---the characteristic features of a hexatic
phase \cite{nelson2}. The system breaks up into crystalline patches of
finite length $\xi_t$ (the translational correlation length) which are
shifted but not rotated with respect to one another, so the range of
orientational correlation $\xi_6$ far exceeds $\xi_t$.  On further
increasing $\Delta$, more defects are created: many monopoles appear
which destroy the orientational order and the system melts
[Fig.~\ref{r9defs}d] (see \cite{nelson3,Rubinstein}).

In the hexatic phase, we find a steep increase in the number of dipoles
(Fig.~\ref{xi6ndef}a) between $\Delta_{\mbox{\scriptsize SH}}$ (the
solid-hexatic transition value for $\Delta$) and
$\Delta_{\mbox{\scriptsize HL}}$ (the hexatic-liquid transition value
for $\Delta$). Also, we detect a gentler increase in the number of
monopoles as the hexatic-liquid transition at
$\Delta=\Delta_{\mbox{\scriptsize HL}}$ is approached. KTHNY
theory~\cite{nelson2} predicts that in the liquid phase, the
orientational correlation function $C_6(r)$ decays exponentially,
$C_6(r) \sim e^{-r/\xi_6}$, with an orientational correlation length
$\xi_6$ that diverges as the liquid-hexatic transition is approached
\cite{nelson2}.  Figure~\ref{xi6ndef}b reveals a rapid increase of
$\xi_6$ as the liquid-hexatic phase boundary is approached from the
liquid side.

Figures~\ref{r1defs}b--d show typical snapshots of the inherent
structure at $\rho = 1.0$, for $\Delta = 0.04$, 0.1 and 0.12. A first
order solid-liquid transition is found at the value
$\Delta_{\mbox{\scriptsize SL}}\approx 0.1$ \cite{srs}, the solid-liquid
transition value for $\Delta$. Our defect analysis shows that for
$\Delta<\Delta_{\mbox{\scriptsize SL}}$, there are free
dislocations, and near the transition these defects line up in
``strings'' to form long chains of large-angle grain boundaries
(Fig.~\ref{r1defs}c). As $\Delta\to\Delta_{\mbox{\scriptsize SL}}$,
these chains percolate, fragmenting the system into micro-crystallites
rotated with respect to each other. Thus the grain boundaries
simultaneously destroy both the translational and the rotational
order. In the theory of grain-boundary-induced melting
\cite{Halperin-Nelson,chui}, the Landau free-energy expansion yields a
first-order solid-liquid transition with the absence of any hexatic
phase.  We find that the formation of defects and the proliferation of
grain boundaries occurs abruptly at $\Delta_{\mbox{\scriptsize
SL}}\approx 0.1$.  Within the resolution of our simulations, we do not
see any hexatic phase \cite{srs}.

In summary, we have seen that size-dispersity induces topological
defects which in turn destroy crystalline order, and that the mechanism
of dispersity-induced melting displays surprising parallels with the
mechanism proposed for the case of temperature-induced melting.
Depending on the value of $\rho$, dispersity-induced melting can be
either a first-order transition or a continuous transition (with an
intervening hexatic phase), and the defect morphologies display
completely different behavior in the two cases.

%% Acknowledgment: We thank A.~Scala for seminal contributions in the
%% initial stages of this work, D.~R.~Nelson for extremely helpful
%% suggestions, and NSF for financial support.

\begin{figure}[htb]
\narrowtext
\caption{
Identification of defects taken from our simulated system. Neutral
($n_i=6$) particles are shown as dots. Solid squares and triangles
denote defects, with $n_i=7$ ($q=+1$) and $n_i=5$ ($q=-1$)
respectively. Neighboring defects form different types of defect
clusters as explained below. (a) The large open square denotes a
positive monopole. It represents a cluster of size one, which is the
positive defect it contains, thus it is a disclination.  The large open
triangle denotes a negative monopole. It represents a cluster of size
three, which consists of two negative and one positive defects.  The
arrows denote dipoles; all but the largest arrow (pointing ``East'')
represent clusters of size two, and hence all but one dipole are
dislocations. The largest arrow denotes a dipole, the vector sum
of two adjacent disclinations forming a cluster of size four with a
non-zero net dipole moment. Finally, the parallelogram contains a
cluster of two $n_i=5$ ($q=-1$) and two $n_i=7$ ($q=+1$) defects with
zero net charge and zero net dipole moment (hence a quadrupole). (b) A
dislocation dipole (solid arrow). Also shown is the corresponding
Burger's vector (dashed arrow), as the vector needed to close an
equilateral (in this case $3\times3$) contour around the
dislocation. The thin solid line marks the extra row which ends at the
dipole. The length of the dipole arrow represents the magnitude of the
dipole $\mbox {\bf P}$,  proportional to the length of the
Burger's vector.}
\label{defsfig}
\end{figure}

\begin{figure}[htb]
\caption{Defects of four different $\rho=0.9$ configurations.
The symbols used for the defects are same as in
Figure~\protect\ref{defsfig}.  (a) A $\Delta=0.04$ configuration; this
configuration is in the solid phase. The magnified portion includes two
dislocations and a cluster of size $10$ (encircled by the dashed line)
with an overall dipole moment.  (b) The inherent structure for the same
$\Delta=0.04$ configuration. Most defects in (a) and (b) form
quadrupoles, and some form dipoles.  Notice that upon forming the
inherent structure of a configuration, the particles with six neighbors
move locally to form an ordered triangular lattice structure. Also many
of the defect clusters, mostly low energy quadrupoles and some dipoles
and monopoles, disappear. There remain only a few defect clusters,
topological defects that cannot be annihilated by local translations
performed in generating the inherent structure; e.g. 
in the magnified region, most of the defects have disappeared; there
remains only one dislocation,  the vector sum of the three
previous dipoles. (c) Inherent structure for $\Delta=0.06$
configuration (in the hexatic phase). The arrows
denote dipoles. Note that free dislocation dipoles appear. (d)
Inherent structure for $\Delta=0.09$ configuration
(in the liquid phase). Many disclination
monopole defects, shown as open triangles and squares, appear. For clarity,
individual defect particles and blobs (including
therefore quadrupoles)  are not shown in c and d. Dipole clusters
are represented by the dipole vector for each cluster.}
\label{r9defs}
\end{figure}

\begin{figure}[htb]
\caption{(a) The number of different defect clusters as a
function of $\Delta$ along the $\rho=0.9$ isochore. There is first a
ten-fold increase in the number of {\it dipoles\/} associated with the
solid-hexatic transition $\Delta_{\mbox{\scriptsize SH}}$. Then a large
increase in the number of {\it monopoles\/} occurs near the
hexatic-liquid transition $\Delta_{\mbox{\scriptsize HL}}$. (b)
Orientational correlation length $\xi_6$ as a function of $\Delta$ along
the same $\rho=0.9$ isochore. Upon approaching the hexatic phase from
the liquid phase (i.e., as $\Delta\to 0.06^+$), $\xi_6$ diverges.}
\label{xi6ndef}
\end{figure}

\begin{figure}[htb]
\caption{Defects of the $\rho=1.0$ system. (a) A $\Delta=0.04$
configuration, which is in the solid phase. The defects are magnified in
the insets. This configuration contains only two dipoles and one low
energy quadrupole. Note that the leftmost and rightmost defects (two
dipoles) cancel, a feature characteristic of the solid phase. (b) The
quadrupole disappears upon generating the inherent structure. 
(c) Inherent structure for a $\Delta=0.095$ configuration which is in
the solid phase near the melting point. (d) Inherent structure
for a $\Delta=0.115$ configuration, which is in the liquid phase. Note
that the defects form lines (grain boundaries) that partition the system
into sub-regions of solid order (micro-crystallites). Unlike the liquid
in Fig.~\protect\ref{r9defs}d, there are few disclination
monopoles. What destroys the solid order is the existence of the grain
boundaries, while in Fig.~\protect\ref{r9defs}d what destroys the solid
order is the existence of both free dislocations and free
disclinations.}
\label{r1defs}
\end{figure}

%%%%%%%%%%%%%%%%%%%%%%%%%%%%%%%%%%%%%%%%%%%%%%%%%%%%%%%%%%%%%%%%%%%%%%%%%%

%\end{document}

%%%%%%%%%%%%%%%%%%%%%%%%%%%%%%%%%%%%%%%%%%%%%%%%%%%%%%%%%%%%%%%%%%%%%%%%%%

\begin{figure}[htb]
\centerline{
\vbox {
    \epsfxsize=7.0cm
    \epsfbox{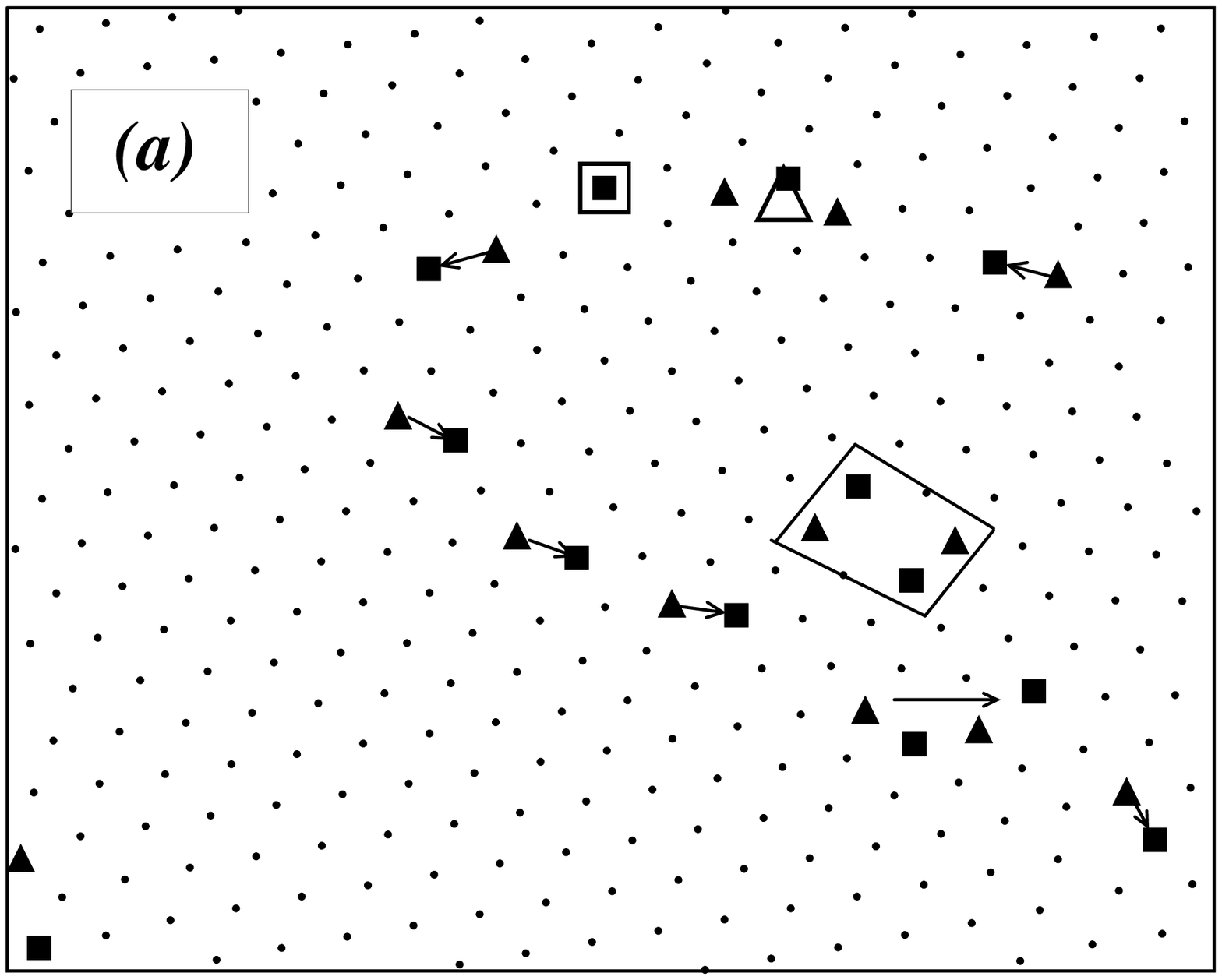}
    }
    }
\centerline{
\vbox {
    \epsfxsize=7.0cm
    \epsfbox{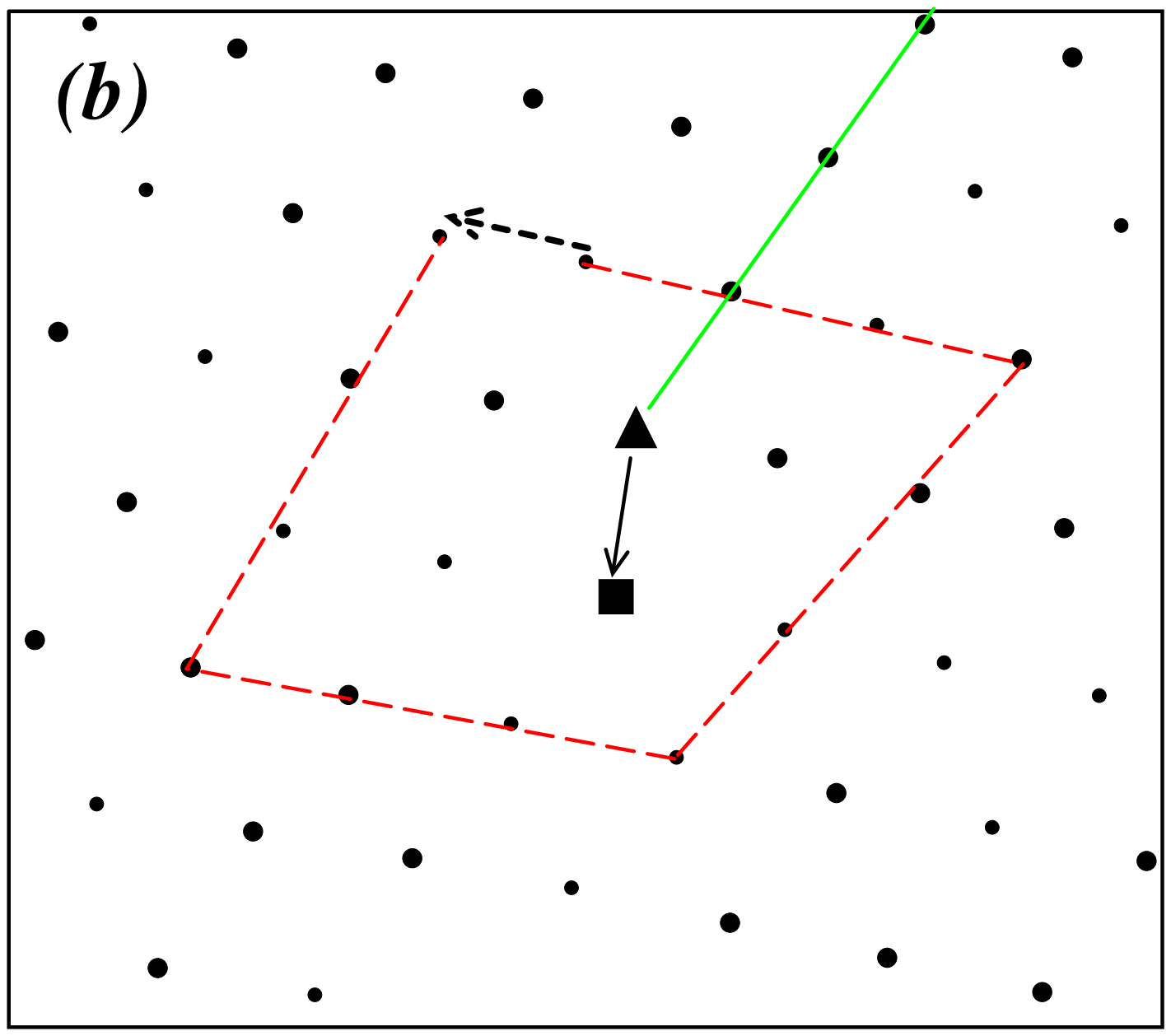}
    }
     }  
\end{figure}

%\newpage 

\begin{figure}[htb]
\narrowtext
\centerline{
\vbox {
    \vspace*{0.1cm} \epsfxsize=7.0cm
\epsfbox{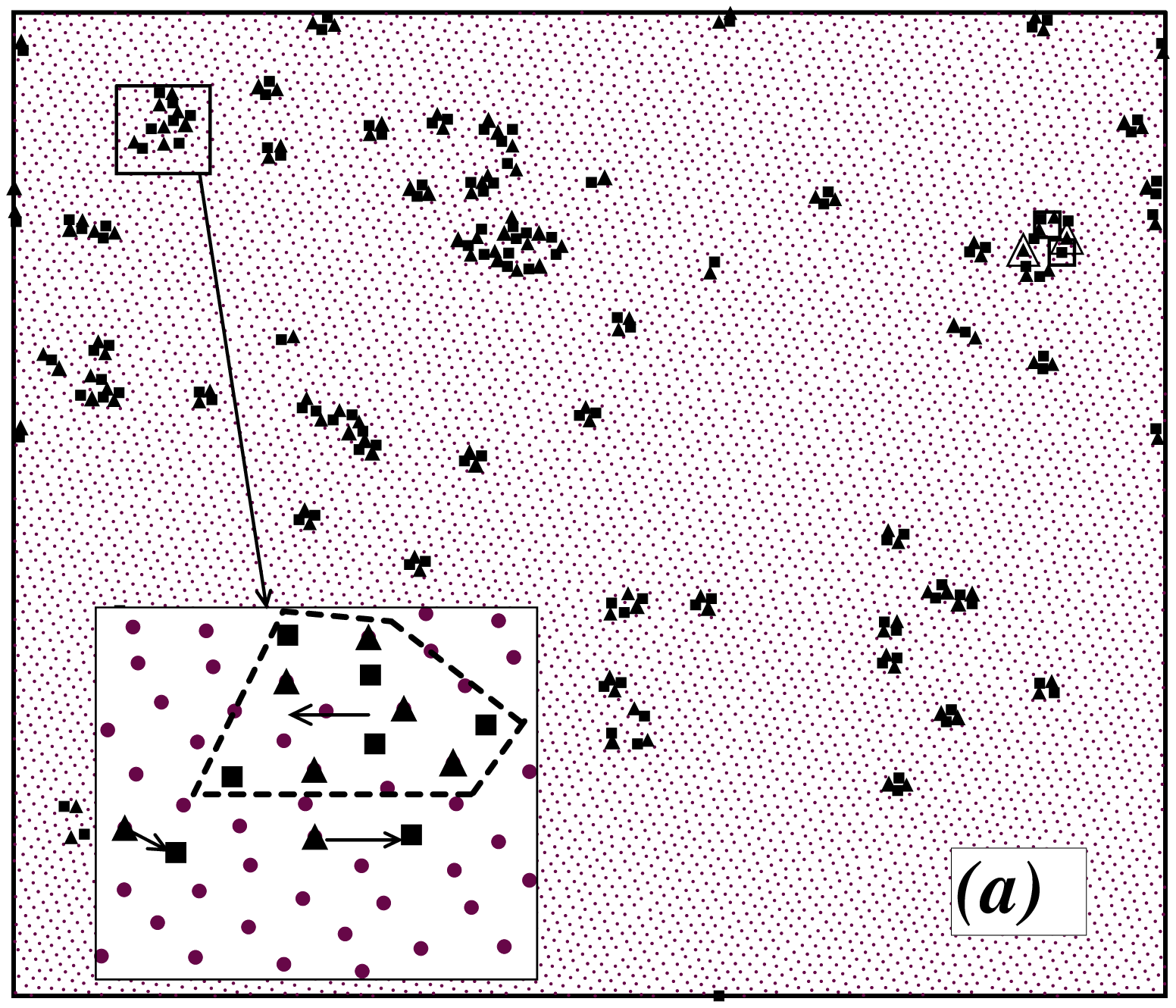} } }
\end{figure}

%\newpage 
\begin{figure}[htb]
\narrowtext
\centerline{
\vbox {
    \vspace*{0.1cm} \epsfxsize=7.0cm
\epsfbox{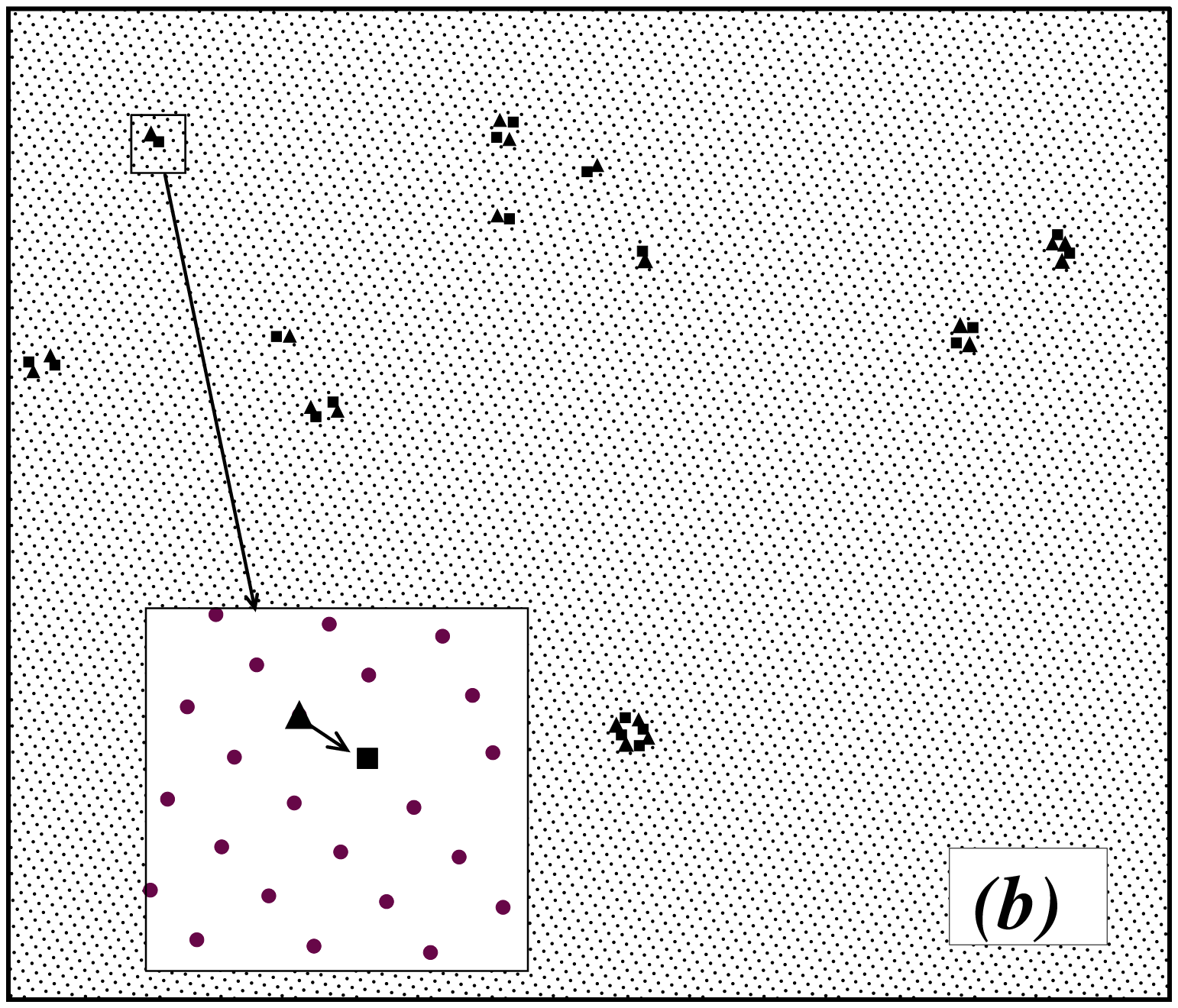} } }
\end{figure}

%\newpage 
\begin{figure}[htb]
\narrowtext
\centerline{
\vbox {
    \vspace*{0.1cm} \epsfxsize=7.0cm
\epsfbox{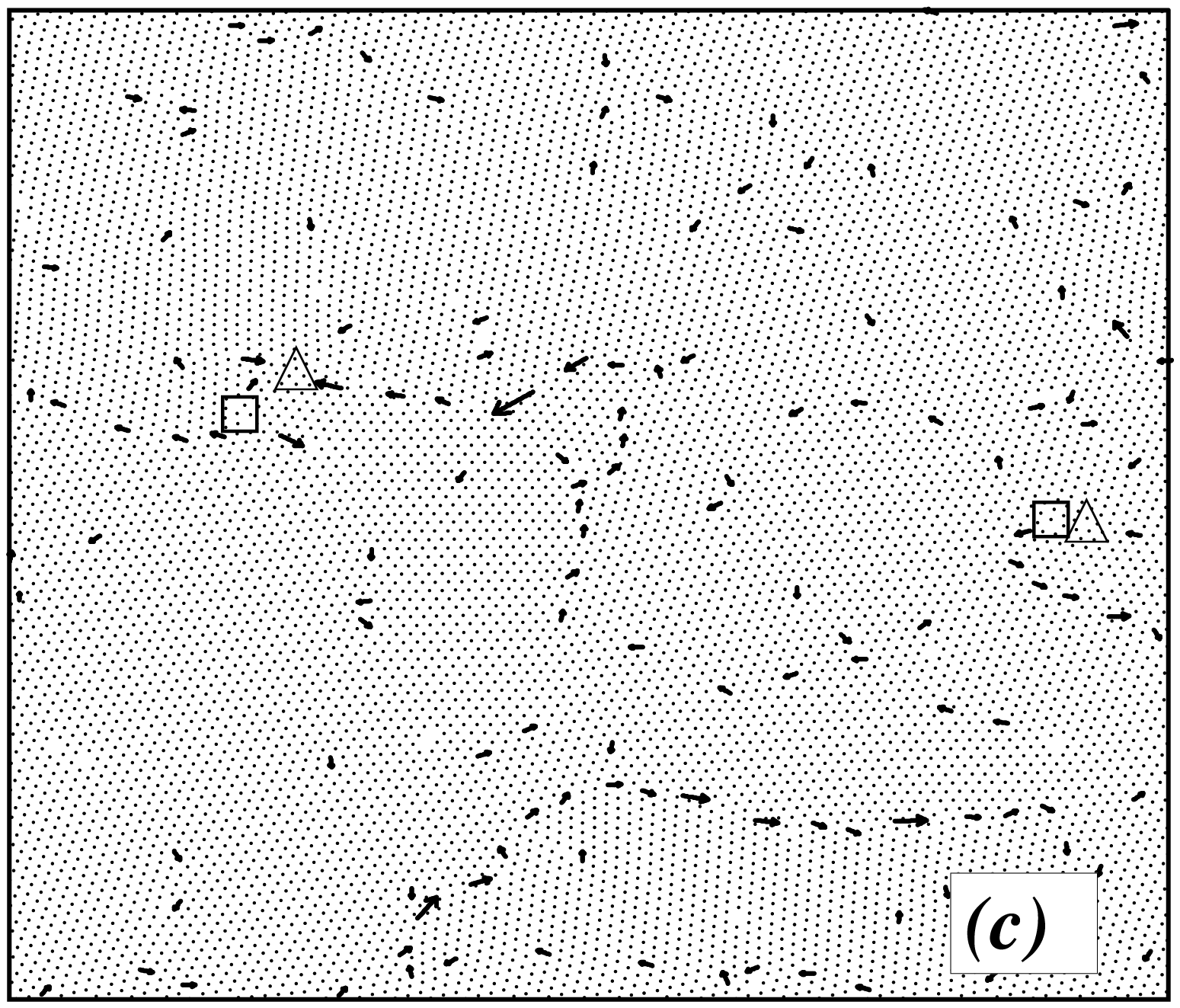} } }
\end{figure}

%\newpage 
\begin{figure}[htb]
\narrowtext
\centerline{
\vbox {
    \vspace*{1cm} \epsfxsize=7.0cm
\epsfbox{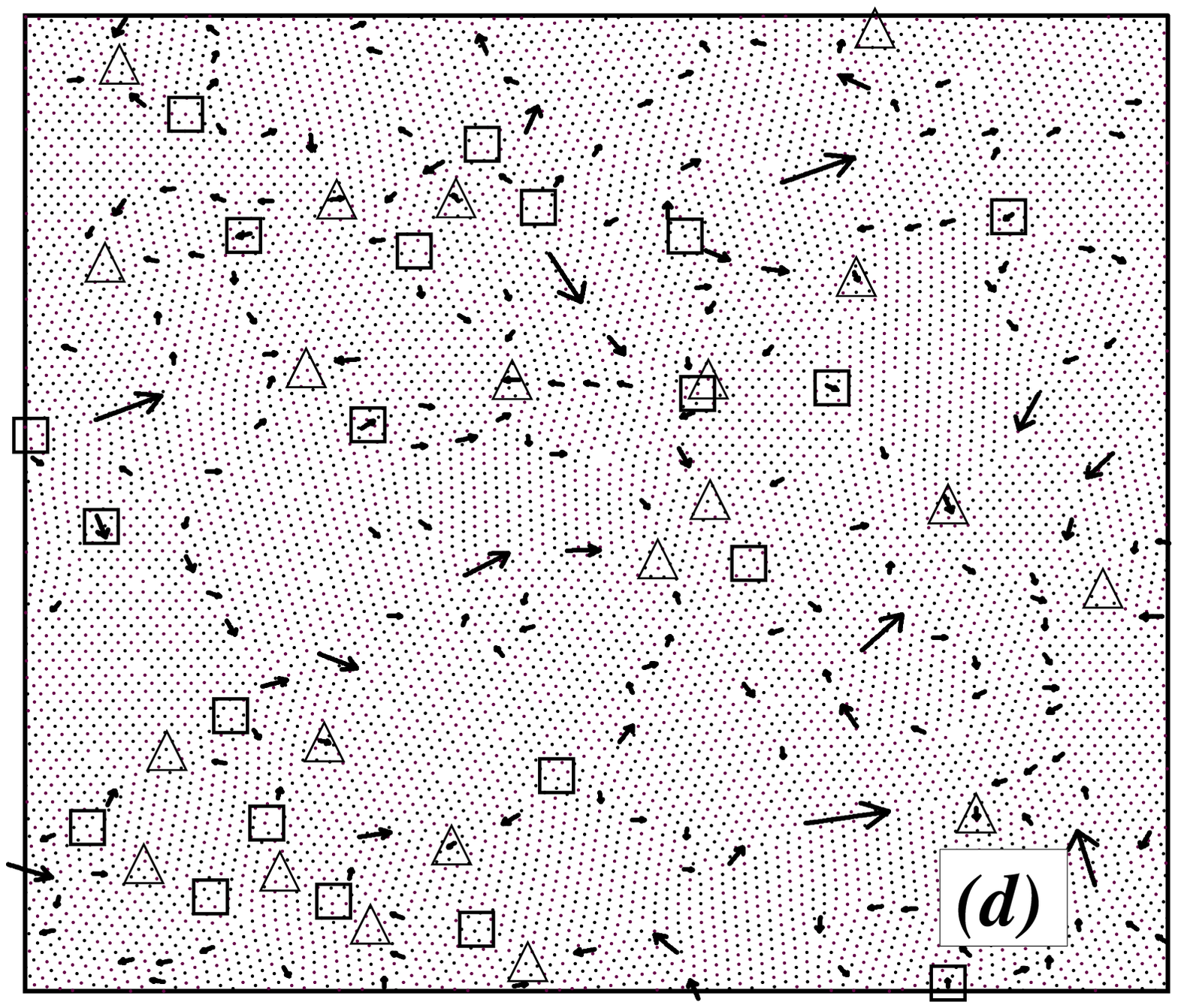} } }
\end{figure}
%\newpage 

\begin{figure}[htb]
\narrowtext
\centerline{
\hbox {
    \epsfxsize=7.0cm \epsfbox{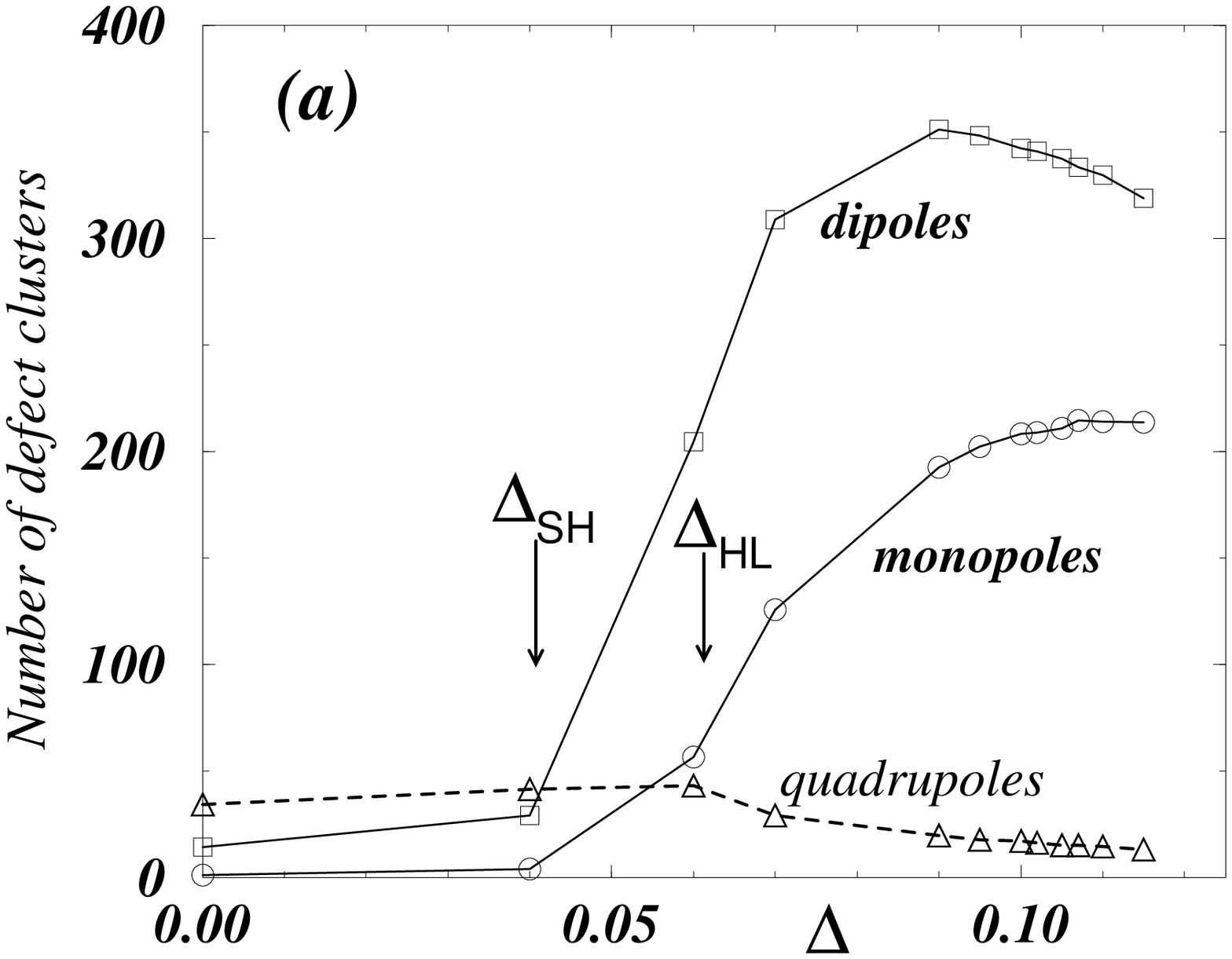} }
}
\end{figure}
%\newpage 
\begin{figure}[htb]
\narrowtext
\centerline{
\hbox {
    \epsfxsize=7.0cm \epsfbox{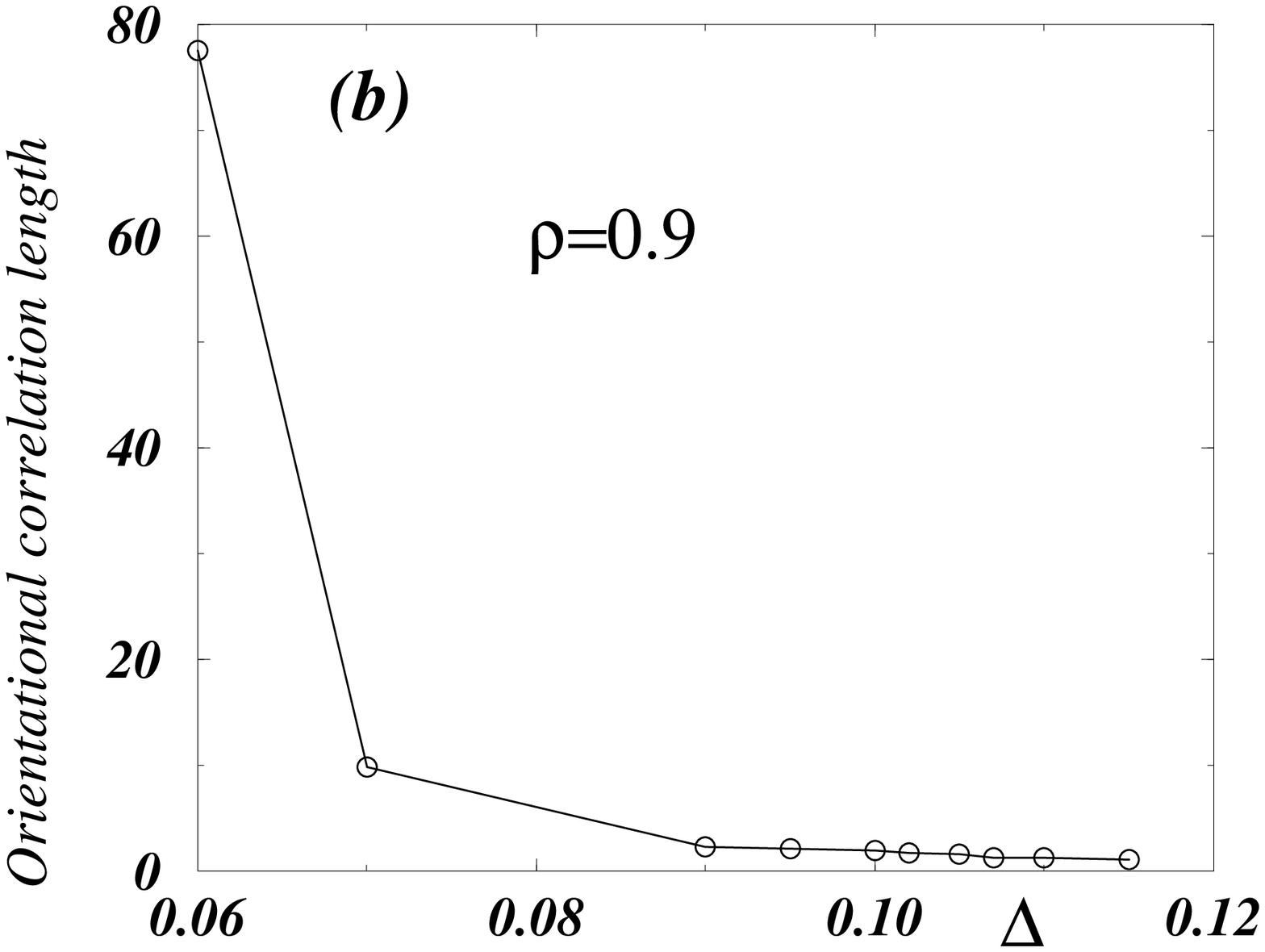} }
}
\end{figure}
%\newpage 

\begin{figure}[htb]
\narrowtext
\centerline{
\hbox {
    \vspace*{1cm} \epsfxsize=7.0cm
\epsfbox{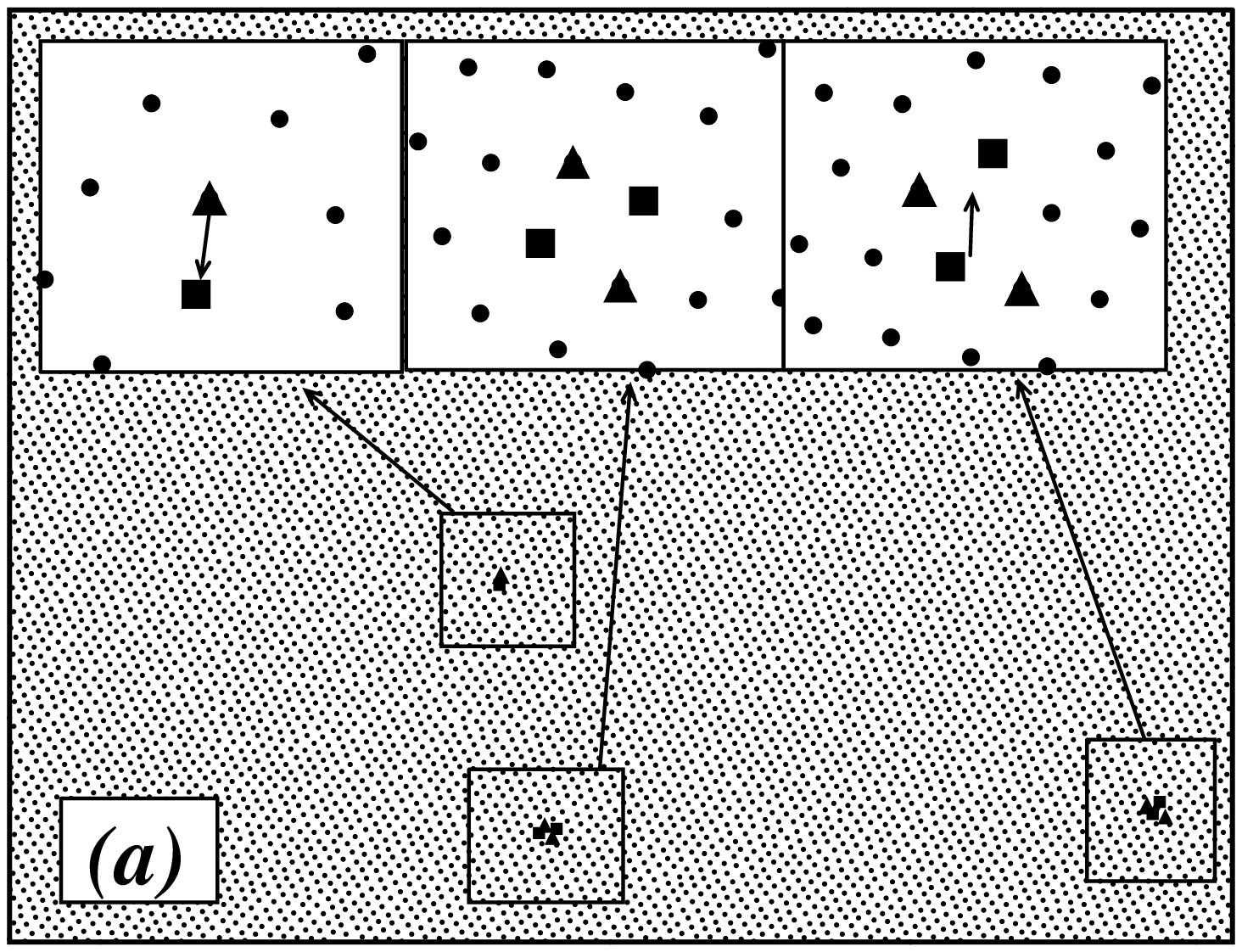} } 	 }
\end{figure}
%\newpage 
\begin{figure}[htb]
\narrowtext
\centerline{
\hbox {
    \vspace*{1cm} \epsfxsize=7.0cm
\epsfbox{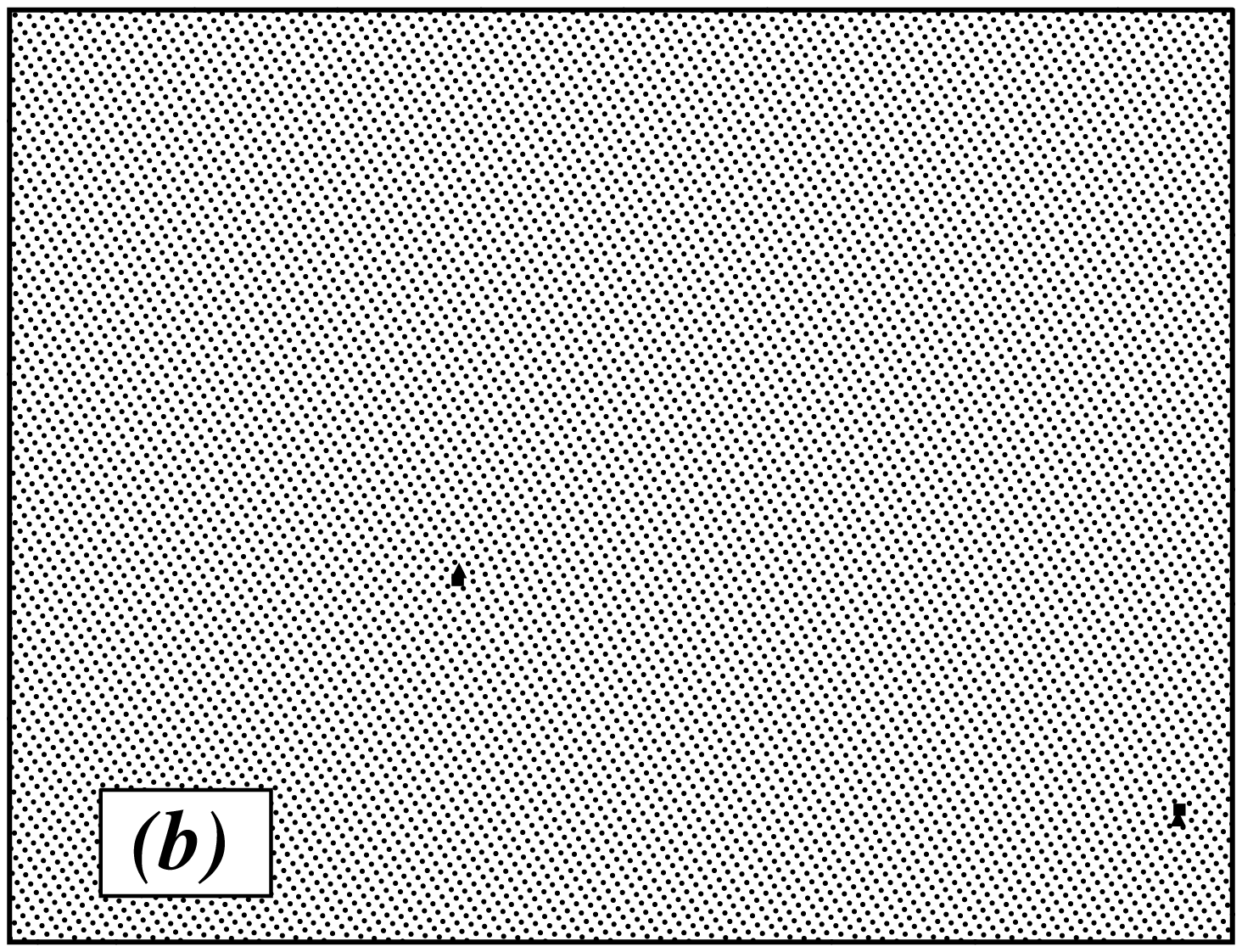} } 	 }
\end{figure}
%\newpage 
\begin{figure}[htb]
\narrowtext
\centerline{
\hbox {
    \vspace*{1cm} \epsfxsize=7.0cm
\epsfbox{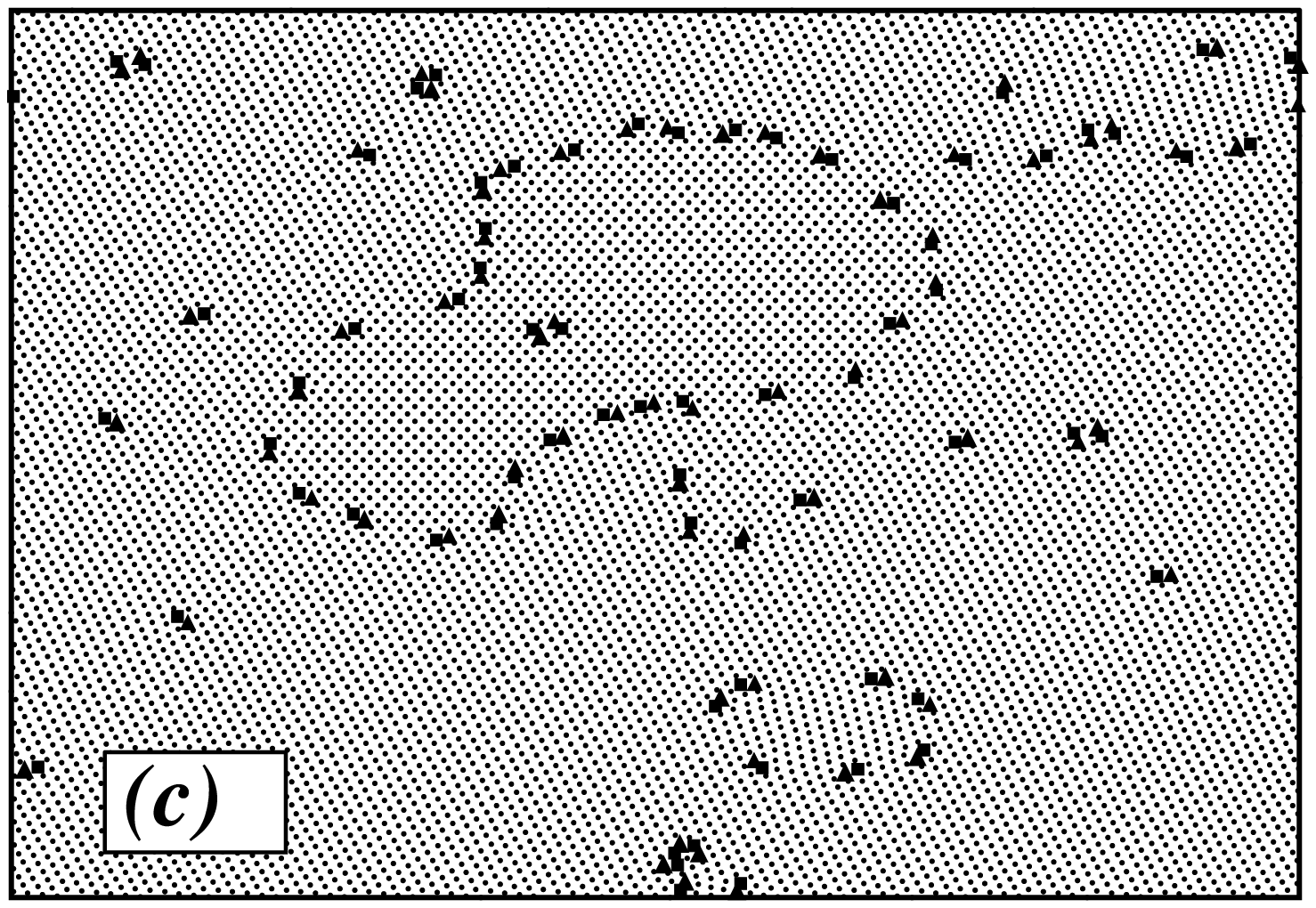} } 	 }
\end{figure}
%\newpage 
\begin{figure}[htb]
\narrowtext
\centerline{
\hbox {
    \vspace*{1cm} \epsfxsize=7.0cm
\epsfbox{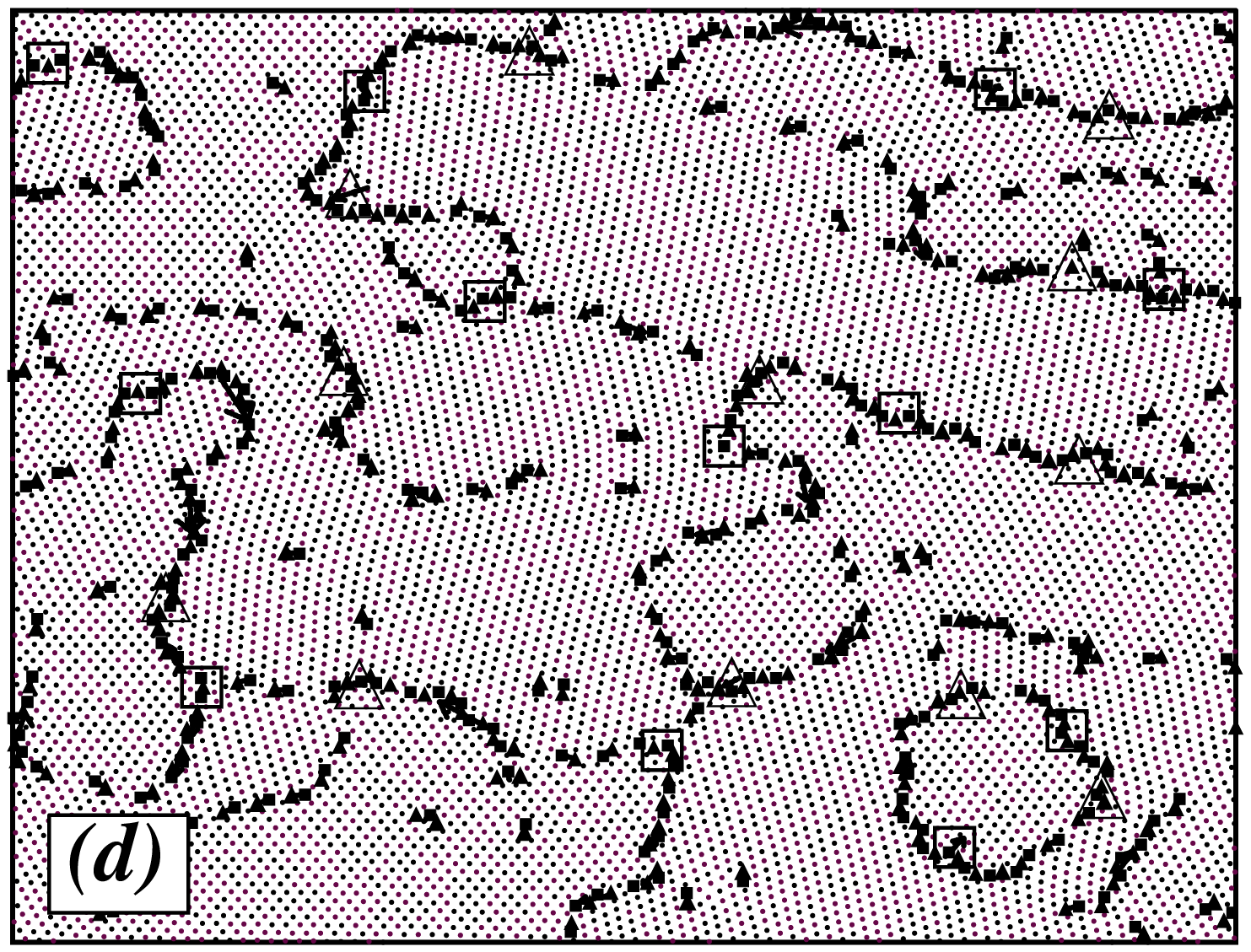} } 	 }
\end{figure}

%\end{multicols}{2}

\end{document}